\documentclass[acmlarge,nonacm]{acmart}

\usepackage{tabularx}

\copyrightyear{2026}
\acmYear{2026}
\setcopyright{cc}
\setcctype{by}
\acmConference[CSCW '26]{The 2026 ACM Conference on Computer-Supported Cooperative Work and Social Media}{October 10--14, 2026}{Salt Lake City, UT, United States of America}
\acmBooktitle{The 2026 ACM Conference on Computer-Supported Cooperative Work and Social Media (CSCW '26), October 10--14, 2026, Salt Lake City, UT, United States of America}

\begin{document}

\title[``I Just Don't Want My Work Being Fed Into The AI Blender'']{``I Just Don't Want My Work Being Fed Into The AI Blender'': Queer Artists on Refusing and Resisting Generative AI}

\author{Jordan Taylor}
\orcid{0000-0002-0896-992X}
\email{jordant@andrew.cmu.edu}
 \affiliation{
   \institution{Carnegie Mellon University}
  \city{Pittsburgh}
  \state{PA}
 \country{USA}
 }

\author{Joel Mire}
\orcid{0009-0003-6230-4229}
\email{jmire@andrew.cmu.edu}
 \affiliation{
   \institution{Carnegie Mellon University}
  \city{Pittsburgh}
  \state{PA}
 \country{USA}
 }

\author{Alicia DeVrio}
\orcid{0000-0002-9912-4198}
\email{adevos@andrew.cmu.edu}
 \affiliation{
   \institution{Carnegie Mellon University}
  \city{Pittsburgh}
  \state{PA}
 \country{USA}
 }

\author{Maarten Sap}
\orcid{0000-0002-0701-4654}
\email{msap2@andrew.cmu.edu}
 \affiliation{
   \institution{Carnegie Mellon University}
  \city{Pittsburgh}
  \state{PA}
 \country{USA}
 }

\author{Haiyi Zhu}
\orcid{0000-0001-7271-9100}
\email{haiyiz@andrew.cmu.edu}
 \affiliation{
   \institution{Carnegie Mellon University}
  \city{Pittsburgh}
  \state{PA}
 \country{USA}
 }

\author{Sarah E. Fox}
\orcid{0000-0002-7888-2598}
\email{sarahf@andrew.cmu.edu}
 \affiliation{
   \institution{Carnegie Mellon University}
  \city{Pittsburgh}
  \state{PA}
 \country{USA}
 }

\begin{abstract}

Art-making is a collective social activity through which queer people engage in political resistance, develop identities, archive queer memory, and form community. However, in recent years, generative AI has disrupted queer artistic communities. Through 15 semi-structured interviews, we examine how queer artists are making sense of the encroachment of GenAI into their art worlds. Our findings surface significant tensions between the relationality of our participants' queer art practices and the perceived anti-relationality of GenAI development and use. We detail how our participants refuse and resist GenAI use and development in response and highlight the limited role our participants saw for GenAI within art-making, such as the queer aesthetic potential of surreal image models. Drawing on queer theory, we discuss how CSCW researchers might support queer artists by refusing dominant AI imaginaries and supporting queer world-building.

\end{abstract}



\maketitle

\section{Introduction}

During the AIDS epidemic of the 1980s and 1990s, art was as a powerful political tool for activism and resistance within LGBTQ+ communities \cite{hawkins1993naming, martin2019art}. Queer people used art to challenge systemic neglect, confront stigmatization, and demand visibility and justice amidst widespread suffering and loss. In this period, Keith Haring made art protesting HIV stigmatization, advocating for safer sex, and remembering those—including eventually himself—who died from AIDS \cite{gruen1992keith}. In one of his last works, Haring left a piece titled ``Unfinished Painting'' incomplete to represent how AIDS interrupts lives. This piece gained renewed prominence in late 2023 when a troll on social media posted a version of the painting that had been completed using generative AI (GenAI), prompting sharp backlash for disrespecting both Haring and others who died from AIDS \cite{ai_haring}. 

Scholar José Esteban Muñoz emphasizes the vital role of queer art in both remembering marginalized pasts and envisioning potential futures \cite{munoz2019cruising}. Muñoz argues that queer aesthetics serve as a dynamic platform for world-building, allowing queer communities to articulate visions of alternative realities. According to Muñoz, queer art embodies a form of utopian potential, creating spaces to explore what could be beyond existing oppressive structures. His scholarship highlights that art is not merely reflective of queer histories, but actively participates in constructing collective imaginaries essential for queer survival and liberation. For Muñoz, relationality is central to queer art, insisting on the ``need for an understanding of queerness as collectivity'' \cite{caserio2006antisocial}. In doing so, Muñoz emphasizes the power of queer art not only to imagine alternative futures but also to refuse oppressive logics in the present, like capitalist alienation.

In alignment with Muñoz's insights, CSCW scholarship emphasizes how queer communities utilize digital spaces to engage in artistic expression as a means of collective identity affirmation and social support. Prior research has demonstrated that participation in online fan communities, such as through writing fan fiction, empowers LGBTQ+ individuals to challenge dominant narratives by creating alternate worlds and possibilities \cite{dym2019coming}. Similarly, Taylor and Bruckman \cite{taylor2024mitigating} have observed that members of online bisexual communities actively perform and explore their identities through shared artistic practices. Furthermore, since queer history is often erased, Riggs et al. \cite{riggs2024designing} argue that art can provide an avenue for archiving and remembering queer experiences.  Such scholarship underscores that online queer communities do not merely circulate art but leverage creative expression as an essential method of community building, resistance, and collective imagination.

Recently, however, the rise of GenAI has significantly disrupted queer artistic communities, raising concerns about the appropriation and commodification of queer creativity. Online fan communities---rich sites of queer creativity and identity work \cite{wang2024counting, taylor2024cruising}---have been particularly affected. The rise of GenAI has led to fears of unauthorized use of original content to train models \cite{fanfic_ai_2025} and has flooded spaces with AI-generated material. In resistance, some artists have engaged in intentional data poisoning campaigns using NSFW content \cite{data_poison_fanfic}, highlighting broader anxieties and collective opposition to GenAI's encroachment. This disruption threatens the very community-driven practices that nurture queer expression and solidarity \cite{dym2019coming, semaan2016transition}.

To understand the impact of GenAI on queer artistic communities, we explore the question: How do queer artists navigate and respond to the growing influence of GenAI? We conducted semi-structured interviews with 15 queer artists, adopting a capacious understanding of queer art that encompasses both the practices of self-identified queer artists and the broader field of queer aesthetics. Echoing \citet{munoz2019cruising}, we find that participants' queer artistic practices are deeply relational and anti-capitalist. Participants juxtaposed the relationality of their art with the perceived anti-relationality of GenAI development and use, such as facilitating worker exploitation. In response, we describe the tactics participants employed to \textit{refuse} GenAI development and use. Finally, we describe the limited queer aesthetic potential our participants saw in GenAI.

The umbrella term ``generative AI'' can refer to many different interaction paradigms \cite{morris2025hci_agi}. This can include the popular prompt-based interfaces that use language to describe desired text or image output, like ChatGPT. However, there are other ways users might interact with generative AI systems, such as the image-to-image interaction used to complete Haring's ``Unfinished Painting'' \cite{ai_haring}. We found that our participants strongly associate the term ``generative AI'' with prompt-based systems, like Midjourney or ChatGPT. As a result, when we discuss generative AI in this work we are referring to the type of prompt-based systems that loomed large in our participants' imaginations.

Drawing from Muñoz's theorization of queer aesthetics \cite{munoz2019cruising}, we discuss how CSCW researchers can support queer art and artists. We contribute to CSCW scholarship on collective resistance through the ways our participants refuse dominant GenAI imaginaries. We also provide implications for the design of online communities to support queer artists and discuss how CSCW and HCI researchers can support queer artistic futures within and beyond GenAI.

\section{Related Work}

This section grounds our work in the literature on queer aesthetics, the intersection of GenAI and art, and queer communities' engagement with technology. First, we introduce José Esteban Muñoz’s theorization of queer aesthetics as a foundational framework for understanding queer art as collective world-making, both within and beyond CSCW research. We then explore recent research on artists’ resistance to GenAI. Lastly, we discuss the complex relationship between queer communities and AI, underscoring the tension between technological inclusion and refusal.

\subsection{Queer Relationality, Refusal, and World-Making}

In the 2000s, there was a heated debate among queer theorists over whether to study queerness through the lens of relationality or anti-relationality, culminating in a 2006 panel on ``The Antisocial Thesis in Queer Theory'' at the Modern Language Association conference \cite{caserio2006antisocial}. In sum, relational theorists highlight the communal nature of queerness \cite{munoz2019cruising}. On the other hand, anti-relational scholars view queerness as an individualist negation of repressive collective obligations, such as biological procreation \cite{edelman2020no_future}. On the panel, José Esteban Muñoz derided anti-relational queer theory as ``the gay white man's last stand'' \cite{caserio2006antisocial}, due to its individualism. Instead, Muñoz insisted on the ``need for an understanding of queerness as collectivity.'' For Muñoz, relationality is crucial for refusing the alienation of capitalism and heteronormativity \cite{caserio2006antisocial}. Elaborating on his theorization of queer relationality, José Esteban Muñoz’s book \textit{Cruising Utopia} explores how queer aesthetics can create utopian alternatives to a cis-heteronormative world governed by ``straight'' logics \cite{munoz2019cruising}. Muñoz conceptualizes utopia not as an abstract future, but as a tangible response to the recognition that ``something is missing'' in the present. Art thus becomes a vehicle for queer world-making, such as the queer intimacy depicted in Frank O'Hara's poetry and ephemeral performances like drag and punk shows.

Relationality has become an important concept in HCI and design research \cite{arturo2024relationality, frauenberger2019entanglement}, emphasizing that people, technologies, and environments are entangled in broader networks of social and material relations rather than isolated interactions. In contrast, anti-relational orientations privilege bounded, individual actors or discrete interactions, often abstracting technologies from the wider systems in which they are embedded. While relational approaches in HCI have drawn attention to issues such as infrastructure \cite{dye2021grano}, maintenance \cite{houston2016values}, and the environment \cite{lu2024unmaking}, much of the field has historically centered individual users and short-term interactions, limiting its engagement with broader relations.

Parallels to queer theory's relationality debate can be seen in recent debates over AI at CSCW. Echoing the aforementioned panel, a 2024 CSCW panel debated ``Is Human-AI Interaction CSCW?'' because CSCW scholarship has historically only included ``computer-mediated collaborations involving at least two people'' \cite{morris2024human}. Although human-AI interaction with chatbots sometimes resembles human-human interaction, critics worry that admitting human-AI interaction dilutes the social aspect of CSCW. Reflecting these concerns, the CSCW 2026 call for submissions states in bold: ``contributions must have a focus on social aspects of technology ... papers whose research contributions are primarily of relevance or benefit to individual users will be considered out of scope'' \cite{CSCW_2026_cfp}. Within this call, one can see an implicit argument that individual human-AI interaction is not ``social.'' In this work, we explore this mismatch between the highly social nature of queer art and the perceived anti-sociality of generative AI.

In both Muñoz's scholarship and CSCW research, collective refusal plays a vital role in resisting capitalist alienation and anti-relationality. Muñoz characterizes the ``queer utopianism'' of art as a ``great refusal of an overarching here and now'' \cite[p. 133]{munoz2019cruising}. In emphasizing its utopian potential, Muñoz suggests that refusal does not merely oppose dominant norms but can also imagine new collective possibilities. A similar throughline can be seen in CSCW scholarship on refusal, non-use, and non-users \cite{baumer2017post, lin2021techniques, satchell2009beyond, sabie2023unmaking, wyatt2003non_user}. For example, one may choose to stop using a social media site to protest the owner's political views \cite{he2023flocking}, a tactic that becomes more powerful when undertaken collectively \cite{vincent2021data}. The binary between user and non-user can also be a fluid site of contestation. Workers, for instance, are increasingly forced to use GenAI or risk losing their jobs \cite{coinbase_firing, jiang2026professional}. Finally, refusing to use harmful technologies is not merely a form of opposition but, like Muñoz argues, a way of making futures otherwise. Janet Vertesi's Opt-Out Project, for instance, encourages people to collectively refuse to use technologies made by Big Tech not only to oppose dominant powers but also because ``there are alternatives out there, better visions of what our lives could be like'' \cite{opt_out}. In this work, we show how queer artists' non-use of GenAI is—rather than simply a choice—a form of collective resistance that requires ongoing work to achieve.

\subsection{Queer Art Worlds as Collective Labor}

Per Muñoz, we understand queer artists as workers and queer world-making as a form of labor. Prior CSCW research supports this view: art-making within bisexual \cite{taylor2024mitigating}, rural queer \cite{biggs2023tiktok}, and fan \cite{dym2019coming} online communities plays a central role in challenging dominant narratives and imagining other ways of living. In these online communities, queer art is not an individual practice but rather a collective endeavor. While these communities share similar commitments to art as a site of queer culture, queer art is not monolithic. Building on sociologist Howard Becker’s concept of “art worlds”---sociotechnical networks of people, practices, and tools that shape how art is made and valued \cite{becker2023art}---we understand queer art as emerging from multiple, overlapping queer art worlds with their own values and aesthetic norms. Our contribution builds on this prior scholarship by examining how queer artists are refusing, resisting, and negotiating the rise of generative AI within these diverse art worlds.

Queer art often challenges dominant aesthetic norms. Camp sensibilities, for instance, subvert dominant aesthetic norms by finding beauty in sincerely ``bad'' or exaggerated art \cite{sontag2018notes}. Avant-garde or transgressive gestures---such as Marcel Duchamp's  \textit{Fountain}, or urinal-as-artpiece---similarly disrupt normative frameworks of artistic legitimacy \cite{camfield1990marcel}. Crucially, queer art is also entangled with memory and collectivity. Because queer history is often excluded from formal archives, queer art becomes a medium for preserving collective memories and embodied histories. Riggs et al. exemplify this in HCI by designing wearable buttons that convey oral histories in ways that evoke queer ephemera \cite{riggs2024designing}. Queer art also operates as a mode of resistance: \textit{The Rupublicans Project}, for example, uses generative AI to satirize anti-LGBTQ+ politicians in drag, while simultaneously raising funds for queer causes \cite{rupublicans}.

Muñoz highlights tensions between the collective labor of queer world-making and the individualistic capitalist ``mandates to labor, toil and sacrifice'' \cite{munoz2019cruising}. He documents how queer performance spaces were shut down in the 1990s and early-2000s by New York City mayors aiming to attract business and tourism, prompting artists to resist through underground performance scenes and activist art in public spaces. Parallel dynamics appear in digital spaces. CSCW researchers have shown how moderation policies designed to appeal to advertisers displace NSFW fan fiction artists from mainstream platforms \cite{fiesler2020moving}, leading some artists to develop their own alternative communities \cite{dym2022building}. These fan-led communities are often non-commercial and participatory by design---practices aligned with queer world-making \cite{fiesler2016archive, dym2022building}. As Bruckman notes in her advocacy for non-profit social media, what is best for members of an online community may not be best for a company's bottom-line \cite{bruckman2022should}.

\subsection{Generative AI, Art, and Capitalist Extraction}

For decades, technology platforms have strongly enforced copyright protection for powerful corporations' art at the expense of independent artists \cite{doctorow2014information}, like YouTube creators \cite{ma2022moderation}, music remixers \cite{lessig2008remix, fiesler2014remixers}, and fan fiction writers \cite{fiesler2015understanding}. At the same time, major companies training GenAI models have heavily relied on text, images, and videos scraped from the internet without regard for independent artists' consent \cite{Reisner_2024, dodge2021documenting, reisner2023these, reisner2023revealed} or ownership claims \cite{goetze2024ai}. Making matters worse, companies in creative industries have begun to use GenAI models to replace or reshape artists' jobs \cite{Merchant_2024}. High-profile artists have also questioned whether GenAI models should—or even \textit{can}—be used to make art \cite{ted_chiang_ai, nick_cave_ai}.

Amid these concerns, artists have found ways to resist and refuse the intrusion of GenAI into their art worlds. These tactics often take the form of data strikes (deleting or withholding one's data) and data poisoning (sharing inaccurate or subversive training data) \cite{vincent2021data}. For instance, fan fiction writers have engaged in data strikes by removing public-facing work \cite{fanfic_ai_2025} and data poisoning by deliberately writing NSFW stories \cite{data_poison_fanfic}. In support of these efforts, computing researchers have developed tools to help textual \cite{agnew2024data} and visual \cite{shan2023glaze, shan2024nightshade} artists protect their work from being used for GenAI training without their consent. Artists have also taken steps beyond the context of data tactics to collectively resist the development and use of GenAI. Subreddit moderators are increasingly creating rules restricting the circulation of AI-generated content in art-related communities \cite{lloyd2024ai_rules}. Artists have also engaged in collective action to resist GenAI in the workplace, such as the 2023 Hollywood film workers strike \cite{halperin2025soulless}. We build on this work by documenting the ways queer artistic communities are resisting and refusing GenAI.

\subsection{Queer AI Research: From Inclusion to Refusal}

Researchers have raised concerns about the impact of GenAI on marginalized communities \cite{jiang2023ai}. For example, it is well established that GenAI models tend to stereotype and under-represent fat \cite{sobey2025thinness}, disabled \cite{mack2024they}, trans \cite{ungless2023stereotypes}, non-Western~\cite{mim2024between, fu2024being_eroded}, and LGBTQ+ communities \cite{gillespie2024generative}. Also, the content moderation policies intended to make GenAI models ``safer'' can inadvertently suppress the work of marginalized artists \cite{mirowski2024robot, taylor2025straightening}. \citet{mirowski2024robot} found that moderation systems ``make LLMs less useful to [comedian] minorities by suppressing content by and about marginalized identities.`` There is also evidence to suggest that some marginalized communities hold more negative views towards AI. For instance, Haimson et al.'s survey of US AI-attitudes found that participants who are non-binary, transgender, disabled, and/or women report significantly more negative attitudes towards AI compared to dominant social groups \cite{haimson2025ai_attitudes}.

At the same time, marginalized artists have found ways to make use of—albeit highly imperfect—GenAI models to support their creative practices. Through a workshop study providing queer artists with GenAI models, \citet{taylor2025straightening} found that queer artists often struggle to use GenAI in their art due to normative values in the design of these technologies, such as social and stylistic biases. Nevertheless, their participants found ways to use GenAI models to support minor aspects of the art-making process, such as drafting client emails. Similarly, Bennett et al. describe how disabled artists leverage GenAI to access new mediums and support peripheral aspects of their creative practice, such as writing, while carefully negotiating their (non)use of GenAI in accordance with their ethical values \cite{bennett2024painting}. Similarly, we examine how queer artists interpret and negotiate GenAI in light of their individual and communal artistic values.

The biases embedded in GenAI models echo a longstanding history of LGBTQ+ people being harmed by the design and deployment of algorithmic systems \cite{taylor2024cruising}. Automated gender recognition systems misgender trans people and uphold harmful notions of gender essentialism \cite{scheuerman2019computers}. Hate speech detection algorithms are often biased against the ways some queer people speak \cite{thiago2021fighting}. Social media algorithms discriminate against queer people, such as via biases in feed recommendations \cite{simpson2021you}, targeted advertising \cite{sampson2023representation}, and content moderation \cite{mayworm2024misgendered}. A group of LGBTQ+ creators famously conducted an audit to prove that YouTube's demonetization algorithm discriminated against their content~\cite{shen2021everyday}. At the same time, it is not always clear what should be done about these biases as queerness being illegible to algorithmic systems can sometimes be beneficial \cite{light2011hci, tran2024making}.

For Muñoz, queerness goes beyond marginalized genders and sexualities. Rather, queerness involves a ``refusal'' of the present ``repressive social order,'' such as cis-heteronormative and capitalist exploitation. As a result, Muñoz is critical of assimilationist LGBTQ+ politics for seeking \textit{inclusion} within systems of oppression, such as the right to serve in the military. Like Muñoz, \citet{hoffmann2021terms} critiques the focus on greater ``inclusion'' in data ethics discourses for overlooking the negative consequences of inclusion in harmful technologies. For example, improving the performance of facial recognition technologies for Black faces may further exacerbate state violence \cite{hassein2017against}. In \textit{Trans Technologies} \cite{haimson2025trans}, Haimson identifies a similar tension between the desire for marginalized communities to be included in mainstream technologies versus create their own. While inclusionism can address concerns in the oppressive here and now, separatism allows marginalized communities to imagine and forge new worlds for themselves. Prior work on the relationship between queer people and GenAI has largely focused on what Haimson dubs ``technological inclusionism,'' such as mitigating social biases \cite{gillespie2024generative}. We build on this work by examining how and why queer artists resist inclusion in the development and use of GenAI.

\section{Methods}
\label{sec:method}

    \begin{table}[h]
    \centering
        \begin{tabular}{>{\centering\arraybackslash} p{0.05\linewidth} | p{0.4\linewidth} | >{\centering\arraybackslash}p{0.2\linewidth}| >{\centering\arraybackslash}p{0.2\linewidth}}

        \textbf{PID} & \centering \textbf{Artistic Practices} & \textbf{GenAI Use \newline (1 = Never, \newline 5 = All the Time)} & \textbf{GenAI Attitudes \newline (1 = Strongly Dislike, 5 = Strongly Like)} \\ \hline
        1 & Video Games, Game Design & Never & Strongly Dislike \\ \hline
        2 & Crafts, Physical 3D Things & Never & Neutral  \\ \hline
        3 & Fan Fiction, Embroidery, Cosplay, Painting  & Rarely & Strongly Dislike \\ \hline
        4 & Musical Theatre & Sometimes & Like  \\ \hline
        5 & Drawing, Sequential Art  & Rarely & Like  \\ \hline
        6 & Visual Art, Fan Fiction, Music. & Rarely & Neutral  \\ \hline
        7 & Fan Art, Fan Fiction, Original Character Design & Never & Strongly Dislike \\ \hline
        8 & Drone/Noise/Glitch Music, 3D Modeling. & Rarely & Neutral  \\ \hline
        9 & Digital Fan Art, Amateur Sculpting  & Never & Strongly Dislike  \\ \hline
        10 & Illustration & Never & Dislike  \\ \hline
        11 & Science-Fiction Writing, Digital Illustration & Rarely & Dislike  \\ \hline
        12 & Crochet, Origami, Lego Building & Never & Strongly Dislike  \\ \hline
        13 & Textiles/Garments, Ceramics, Printmaking & Never & Strongly Dislike \\ \hline
        14 & Cartoons, Non-Representational Drawing & Never & Dislike  \\ \hline
        15 & Social Practice, Sculpture, VR & Rarely & Neutral  \\ \hline
        
        \end{tabular}
    
    \caption{Participant demographics detailing their individual artistic practices, how often they use GenAI (1 = Never, 5 = All the Time) and how they feel toward GenAI (1 = Strongly Dislike, 5 = Strongly Like).}
    
    \label{tab:participant_demo}
    \end{table}

    In this work, we conducted semi-structured interviews with 15 artists we recruited through a web form shared by members of the research team on Reddit and Twitter. Our study was approved by our university's institutional review board. We required our interviewees to (1) be at least 18 years old, (2) live in the United States of America, and (3) identify as ``queer artists.'' In our recruitment form, we asked participants to describe their artistic practice. Our participants engaged in a variety of art forms, such as writing, visual art, textile art, sculpture, music, and video game design. About half our participants mentioned engaging in fan communities (P3, P6, P7, P9, P11, P14) through either fan fiction or fan art. Through our interviews, we found that only one of our participants primarily made a living with their art (P1). Other participants sold their art in more one-off ways (P3, P6, P7, P14, P15), such as selling fan art at conventions (P7). We required participants to rate the following on a five-point scale: how often they use GenAI (1 = Never, 5 = All the Time) and how they feel toward GenAI (1 = Strongly Dislike, 5 = Strongly Like). Our participants identified as infrequent GenAI users (mean = 1.53, median = 1 [Never]) and tended to hold negative attitudes toward GenAI (mean = 2.13, median = 2 [Dislike]). There was greater variation in how our participants felt toward GenAI (std = 1.13) than in how often they use GenAI (std = 0.64). More information can be found in Table \ref{tab:participant_demo}. Our participants were also given the option of sharing additional demographic information. Of those who provided their age, they ranged from 20 to 33 years old, with a median age of 26. A majority of our participants identified as white, five as Asian, and two as Hispanic/Latina. Due to our inclusion criteria, all participants identified as queer artists. Most of our participants~(10) identified as trans, non-binary, genderqueer, or gender non-conforming. Four participants identified as disabled.

    We did not set out to recruit participants who were particularly critical of GenAI. In fact, our recruitment materials explicitly stated that we were looking to understand queer artists' ``attitudes (positive or negative)'' toward GenAI, in an attempt to appeal to both proponents and opponents. Nevertheless, those who responded to our recruitment call tended to hold more negative or skeptical attitudes. While we make no claim that our participants are representative of all queer artists, our participants' negative attitudes toward GenAI might be unsurprising as transgender, non-binary and disabled people in the US tend to view AI much more negatively than members of dominant social groups \cite{haimson2025ai_attitudes}. After finding that respondents to our recruitment surveys tended to view GenAI disproportionately negatively, we decided to investigate the source of this skepticism and non-use in greater detail. That said, future research should give specific attention to the subset of queer artists who are GenAI proponents.
    
    Each semi-structured interview lasted approximately one hour, and each participant was compensated with a \$25 Amazon Gift Card. The first author led each interview, which took place between March and May of 2024. Our questions focused on understanding the relationship between our participants' identities and their art, their past experiences with and feelings toward GenAI, and how GenAI has been impacting their artistic communities. However, our questions narrowed slightly over time. Based on nascent themes from our earlier interviews, we modified our questions over time to theoretically sample for a greater understanding of values associated with art and GenAI \cite{corbin2014basics}.
    
    First, we transcribed the interviews using Zoom's audio transcription feature. Then, the first and second authors listened to each recording to manually correct the automated transcripts. After transcribing each interview, the first and second authors engaged in open coding \cite{corbin2014basics}. This open coding process lasted approximately one month. While coding, the first and second authors wrote memos and discussed initial patterns in weekly meetings. Then, we imported each of our open codes onto a digital whiteboarding tool to perform axial coding. At this stage, we identified numerous dialectical tensions between our participants' perceptions of art and GenAI, such as ``art-making as a process'' versus ``generative AI as a shortcut.'' This led us to locate the source of our participants' antagonism toward GenAI as resulting from value tensions in the production of art. These tensions are reflected in Section \ref{sec:harm}. In light of these tensions, we noticed our participants engaging in strategies to resist GenAI (Section \ref{sec:strategy}). Finally, we found that our participants' attitudes towards GenAI more nuanced than a simple good/bad binary; rather, their views depended on whether models were being used or developed in ways that align with their individual and community values. In Section \ref{sec:benefit}, we describe the—albeit limited—potential our participants saw in GenAI to support art-making.

    Our methodology has been shaped by the positionalities of our research team. Multiple authors identify as members of the LGBTQ+ community and all authors identify as white and/or East Asian. Our interpretations were also shaped by our academic disciplines as design, CSCW, and NLP researchers as well as our experiences practicing and appreciating musical, visual, written, and textile arts. Our team is located at a university in the United States of America, which restricted our ability to recruit participants living outside the United States of America.

\section{Findings}

In this work, we interpret the experiences of queer artists through Muñoz's queer theories \cite{munoz2019cruising}. For Muñoz, queer art is about far more than art made by queer people. Queer art is a vehicle for collective world-making and a refusal of capitalist alienation in the present. It follows that not all art made by an LGBTQ+ person is necessarily ``queer art'' in the queer theoretic sense when one's work upholds systems of oppression \cite{duggan2002new}. That is to say, AI-generated images shared by Sam Altman, the self-declared ``techno-capitalist'' billionaire CEO of OpenAI, are not ``queer art'' solely by virtue of his gay identity \cite{sam_altman_bio}. In this section, we first describe the queer, relational, anti-capitalist nature of our participants' artistic practices. Next, we juxtapose the relationality of our participants' art with the perceived anti-relationality of GenAI development and use. Then, we describe how our participants resist these anti-social GenAI artistic practices. Finally, we elaborate on the nuanced potential our participants saw for GenAI to support queer art making.

\subsection{Queer Art is Highly Relational}

    The utopian potential of queer art can be seen in the ways our participants used art to craft collective queer worlds. P14 is an artist whose work is ``heavily influenced by the furry fandom.'' He uses art to realize: ``desires that don't exist in my life, that don't exist in the world, or might be impossible to exist in the world.'' Through his art, P14 is able to explore his identity and re-examine childhood experiences of ``alienation.'' Similarly, P3 explained: ``It was through fan fiction, fan activities and fan art that I really understood that I was queer and saw positive expressions of queer love and joy in a relationship.'' Moreover, through her own queer fan fiction, P3 has supported others by participating in ``quite a few charity zines'' to raise money for LGBTQ+ and disability causes. P1, a queer game developer of color, integrates characters with various skin tones and LGBTQ+ identities into her games but ``never has it be a central thing of their character.'' In doing so, P1 materializes a world wherein identities ``show up the way [she] would want them to show up in more general media: just like happenstance.'' Through their art, our participants make queer futures in the present.

    Queerness can also be seen in our participants' art through community building. Neither P12 nor P15 described their art as directly related to queer identity, yet both described their artistic practices as ``fruity.'' P12 notes how in the USA men are not ``taught to show emotion,'' ``appreciate their friends,'' or crochet. Therefore, he views ``gifting [his crochet art] to friends'' as a queer act. For P12, his art is less about crocheting a ``rainbow'' to \textit{represent} queerness than the ``act of \textit{doing} [emphasis added]'' queerness by making textile art as a gift for others. Likewise, P15's sculpture and social practice art is not intentionally queer, but P15 still hopes their art will connect with those ``who are of a similar identity.'' In other words, queer art does not depend solely on queer subjects within art, but also queerness as a way of forming community. 

    Although our participants made art in a variety mediums—including writing, visual art, textile art, sculpture, music, and video game design—they shared a similar relational orientation. For our participants, art was rarely about making artifacts in-and-of-themselves but rather the collectivity art can foster. This can be seen in the importance of online communities, such as Twitter (P3, P6, P7), Reddit (P1, P8, P12), and Discord (P6, P9), as well as in-person gatherings, like fan conventions (P3, P7), in our participants' art worlds. The relational, anti-capitalist nature of queer art can also be seen in how often our participants make art as a gift (P3, P5, P9, P12, P13). P9, a fan artist, will ``make people art as gifts'' because they like making ``something to bring people happiness.'' Likewise, P13 gives their friends pottery to remind their friends ``[P13] loves me,'' saying ``I like that about art.'' As we will describe below, our participants sharply contrasted the relationality of their artistic practices with the perceived anti-relationality of GenAI.

\subsection{Generative AI is Perceived as Anti-Relational}
\label{sec:harm}

As described above, our participants maintained artistic practices enmeshed in webs of relationality. Rather than being motivated by profit maximization, our participants used art to care for others and build alternative worlds. However, our participants felt that generative AI models are often developed and used in ways that contradict this collectivity, such as training GenAI models on art without artists' consent. Additionally, our participants worried that GenAI models facilitate the exploitation of artistic workers, violating the anti-capitalist ethos of queer art.

\subsubsection{Participants believed that GenAI developers do not care about the artists whose work they train on}

    Our participants expressed frustration that GenAI models were built from art without creators' consent, describing GenAI development as ``stealing'' (P1), ``theft'' (P6), ``scummy'' (P7), ``nasty'' (P9), and an ``invasion of privacy'' (P13). As P11 explained, ``Artists are not super happy about generative AI because they were not being properly taken care of or consulted and no one asked their permission.'' Here, the use of the word \textit{care} highlights a critical distinction between the way participants treat others versus the way they feel treated by GenAI developers. P11 feels like GenAI developers \textit{do not care} about the artists who produced the art they scrape. Drawing on their own experiences, P13 worried: 

    \begin{quote}
        ``I wanna be able to be certain that Microsoft isn't pulling from a Tumblr blog I've forgotten about from 2013 with pictures of my art on it or something. I just don't want my work being fed into the AI blender ... You cannot be certain that if you opt out that your choice is actually being respected.'' (P13)
    \end{quote}

    In doing so, P13 contrasts the relationality of art shared to an imagined audience on social media with the anonymous agglomeration of art to train GenAI models. Not only did P13 fear having their work ``fed into the AI blender'' without consent, their concerns were exacerbated by the inability to opt-out. In light of the damage already done by generative image model developers, adding an option for artists to opt-out would not resolve P13's concerns because they do not trust developers to respect their choices.

    In addition to concerns about their own art being used to train GenAI models without their consent, our participants compared non-consensual scraping to existing taboos within their artistic communities. P9 explained how non-consensual scraping violates norms within the visual fan art community against ``tracing,'' or directly copying another fan artists' work without attribution or consent:

    \begin{quote}
         ``Having a generative AI that is trained on a consensual dataset is more ethical to me than one trained on stolen data, just like `tracing' versus taking a publicly available base and using it. Because the base creator wanted people to use their stuff while the tracer or the person you traced probably did not.'' (P9)
    \end{quote}

    In distinguishing between tracing one's art and using a ``publicly available base,'' we can see that the acceptability of how one uses art is also tied to the artists' expectation of how their work will be used. While tracing a work that is not shared to be used in this way is unacceptable, those creating base templates likely expect their work to be traced.
    
    At the same time, improved data consent was not necessarily perceived as a panacea for the ethical concerns regarding GenAI. P7 felt that seeking data consent would make GenAI feel ``a lot more ethical.'' However, she still worried that the resulting models might be used in ways that artists might not desire, such as visual art that is ``not safe for work'' or ``derogatory to other people.'' Of course, P7 worrying that one's art could help GenAI models create images ``derogatory to other people'' requires caring about ``other people.'' In sum, our participants' concerns about non-consensual scraping are about not only individual artistic agency but also respect for others.

\subsubsection{Participants believed that using GenAI can alienate art from artists}

For our participants, art was inextricable from the deeply human process through which it is made. For this reason, participants often used the metaphor of having, lacking, or being filled with a ``soul'' (P1, P2, P3, P7, P10) to explain their feelings toward GenAI. P12 told the story of a textile artist who made a blanket for their baby, saying he enjoys ``the humanity part'' of art: ``I know people say you can separate the art from the artist, but I disagree.'' P9 similarly felt that GenAI models alienate art from artists: ``There is an argument for why [using GenAI] shouldn't be disappointing being that 'Oh, look at the product and stuff.' But that sort of approach to making stuff leads to a less empathetic more miserable world, one more susceptible to consumerism and working for our Lord and saviors Amazon and being exploited until the end of time. We need to value people around us more than we value products.'' Here, P9 draws a sharp distinction between their art and those they imagine using GenAI. While P9 ``values the people around us,'' they cast AI-generated art as a capitalist ``product'' alienated from social relations.

Participants placed a similar emphasis on the embodied, material history of art, evident in the comparisons they made between GenAI and other technological mediation. P6 explained that knowing GenAI was used to make a piece ``affects the way I feel about the piece in the same way that any other kind of tool would.'' To explain, P6 imagined seeing a photorealistic image online: ``If I see a caption that it was a picture, I'd be like, `Oh, cool! That's a cool picture.' But if underneath I see that this is a painting done on a canvas with acrylic, I'm like `Mind blown? That's amazing!''' Likewise, the embroiderer P3 considered both hand embroidery and machine embroidery to be ``just two different ways of approaching a piece of art,'' even though the latter is often faster and cheaper to make than the former. However, unlike with GenAI, P3 felt that ``people who do machine embroidery aren't trying to pass it off'' as hand embroidery.

Participants reproached the the undisclosed use of GenAI for perpetuating artistic alienation. P1 explained: ``My feelings towards [GenAI] are pretty negative right now. They mainly seem to be used to — I mean to be blunt — scam people.'' Similarly, P7 worried: ``Just the fact that I would be able to find out that something's AI. It gets upsetting because it's becoming more and more like deception.'' In response, participants advocated for greater disclosure. P4 thought ``if you didn't do all the work, you should probably mention that you used AI.'' Likewise, P2 encouraged artists to ``just be open about what you're doing with [GenAI].'' Some participants (P9, P10, P11) wished AI-generated images were labeled when shared on social media. P11 was glad to see that visual and written art contest submission forms increasingly ``ask you to click the box if this is AI-generated.'' P11 went on to say that without these ``labels'' you ``cannot track it down'' if ``original artists claim that they have had their copyright violated.'' They also thought that better detection tools might help address the negative consequences of AI-generated spam, which has led some sci-fi magazines where they submit to pause open calls.

\subsubsection{Participants believed that GenAI models facilitate the exploitation of artistic workers}

    Our participants critiqued those who use GenAI to cut costs. This can be seen in numerous participants describing GenAI as feeling ``cheap'' (P1, P3, P5, P11, P12). P1 explained: ``There's always going to be people who take stuff that's meant to be tools or minor shortcuts, but then they use \textit{all} the shortcuts.'' P1 likened GenAI to ``asset flipping,'' when ``game developers buy a bunch of pre-made assets, put them into a really quick game and then sell it for profit.'' P14 raised similar concerns: ``I can tell like, `oh, they didn't budget for hiring an illustrator to make a bespoke image for this because they had a Midjourney subscription.' In those sorts of cases, that's just laziness. That's not necessarily an AI problem. That's a laziness problem. That is a lack of imagination problem.'' While our participants viewed art as inextricable from artists, participants worried that GenAI makes it easier to extract profits from art by cutting artists out of the art-making process.

    Although most of our participants did not make money from their art, they often worried about the labor impacts of GenAI on other artists. For example, P10's ``main concern'' regarding GenAI was that ``people are using [GenAI] at the expense of working artists, writers, actors, etc.'' That said, our participants did not necessarily believe that GenAI could \textit{truly} replace artists' jobs. P14 characterized employers' present use of GenAI as ``a violence against those artists who now are finding themselves out of work because the people who would hire them falsely believe that AI is capable of replacing that type of work.'' P6 worried managers will use GenAI to pressure artists: ``The expectation is no longer going to be `You have a week to do this' now it is `Okay you have 2 days to do this because we've given you this new tool, which means that you should be able to do it faster.' '' P1 echoed:

    \begin{quote}
        ``There's this inherent joy in art of facing a problem and coming up with a solution on your own. That's why I think, for a lot of game developers, AI doesn't seem that enticing as opposed to CEOs and more business oriented people who want to cut as many game developers out of the development process. This probably contributes to the stigma. It's literally the machines taking dream jobs away.'' (P1)
    \end{quote}

    Because our participants worried about GenAI being used to financially harm artists they often distinguished between those using GenAI to make money versus those using it for personal use. In particular, our participants typically perceived the latter to be more ethical. As P3 explained: ``I feel like using it for your own purposes is fine, perhaps, but if it's taking away people's jobs, or if you're making a profit off of it that I feel way less comfortable with.'' P11's attitudes toward GenAI use depended on whether the art was ``personal versus public'' as well as whether a piece of art is made ``for profit.'' P11 had few qualms with private usage or public usage for non-profit purposes. However, she thought that when GenAI is used for profit there is a need for ``more regulation and transparency.''

    Echoing the anti-capitalist values of Muñoz's queer utopian logics \cite{munoz2019cruising}, participants routinely located their concerns with GenAI not entirely within the technology itself but rather the capitalist system. Rather than debating if GenAI should be used at all, P6 wished artists would focus more directly on ``workers' rights in creative industries,'' such as ``seeking to regulate the ways that we use [GenAI] or regulate the treatment of workers using it.'' Similarly, P8 focused on critiques of capitalism rather than GenAI: 

    \begin{quote}
        ``I just feel like if people didn't have to rely on their art to make money and support themselves that would be a better world. I feel like we wouldn't be having these conversations about the fear of being replaced. It's only because there's this need to extract profit from art.'' (P8)
    \end{quote}

    As P15 summarized, ``being a creative person under capitalism is hard'' because ``good art'' is not necessarily the most ``financially successful art.'' In other words, our participants' anti-capitalist values are in tension with the material realities of making art within a capitalist system. Our participants felt that GenAI threatens artistic labor because of the ways in which it can be used to make art more profitable, often at the expense of quality.

\subsection{Refusing GenAI as World Building}
\label{sec:strategy}

As described above, our participants' queer art was deeply relational. In turn, they reproached the perceived anti-relational nature of GenAI development and usage practices. Much like Muñoz describes queer art as a ``refusal'' of the present ``repressive social order,'' our participants tried to refuse the ``straight'' logics of GenAI in their art worlds. First, our participants tried to refuse participating in GenAI training by limiting the visibility of their art and refusing to use GenAI built on non-consensually scraped data to stand in solidarity with other artists. Moreover, our participants tried to detect and call out undisclosed GenAI use to resist the use of GenAI within their artistic communities. In much the same way that a queer punk show can provide a joyful reprieve from heteronormative life \cite{halberstam2003queer_temp}, our participants' refusal of GenAI is not only a negation of a repressive present but an act of queer world-building. Amid the increasing omnipresence of GenAI, trying to maintaining art worlds without GenAI is radical. At the same time, this refusal can come at a cost, like artists now feeling the need to prove their work is ``real.''

\subsubsection{Non-Use}

    In light of the ethical concerns described above, our participants often refused to use GenAI as an act of resistance. When asked why she had not used GenAI, P10 explained that it was a ``deliberate'' decision because the models are built using ``people's existing work without that consent.'' Likewise, P3 felt ``guilty'' for having used a generative language model once to explain a technical concept to her. When asked why, P3 explained that using ChatGPT felt like ``almost a type of betrayal to my values and how much I value the work of writers and also artists'' because ``AI generated art has taken away a lot of jobs'' and the model is ``trained without people's consent.'' Similarly, P15 has not used GenAI in their art because they ``really value artistic integrity.'' In sum, our participants stood in solidarity with other artists and enacted their values by choosing not to use GenAI. In a situation where individual artists have limited abilities to resist the major corporations developing GenAI models, non-use allows participants to refuse the normalization of GenAI and build collective solidary with other artists.

    Our participants also described numerous ways they and members of their artistic communities are resisting those building AI models without artists' consent. For example, P13 chose to limit the visibility of their visual art: ``I don't put my art online anymore. I don't want [fast fashion brand] stealing it. I don't want AI combing through it to take chunks of it and cut it apart and use it for other things I don't want. I don't want any of that. So I purely share my art person to person.'' In an attempt to regain agency, P3 saw fan fiction writers place `do not scrape' requests at the top of stories: ``They'll put something up top that says like `Please don't use this for AI training' ... not that I think that [the requests] would actually stop anybody.'' Within this caveat, P3 acknowledges the limitations of trying to avoid having one's art used to train GenAI models. One could choose not to share their art online altogether, like P13, but this severely limits who one can share their art with. If one chooses to share one's art online, they risk having their art scraped (even if they clearly expresses a desire to opt-out).

    For some participants, GenAI (non)use was more of a spectrum than a clear binary. P11, for instance, said she would never use GenAI to write her queer science fiction stories. However, she has used GenAI to help with ``preparation'' (e.g., information seeking) and ``post-production'' (e.g., editing pitch letters or synopses for writing competitions), tasks which feel peripheral to her writing practice. Participants often made sense of AI acceptability through public controversies over artists using AI, such as a a YouTuber criticized for using GenAI to create visual effects (P7) and animators of a recent film accused of using GenAI (P3, P9, P10). In justifying animators using GenAI, P10 thought that if GenAI ``can make a job just a bit easier, then I think that's a good application.'' Nevertheless, participants described AI use as highly stigmatized, a stigma enforced through call-outs. Below, we describe the implications of these AI-use call-outs in greater detail.

\subsubsection{Detection and Call-Outs}

    Due to the lack of transparency over GenAI use, participants developed and deployed their own strategies to try to detect AI-generated images. For example, our participants described looking for idiosyncrasies in people's ``hands'' within visual art (P3, P7, P9), such as counting the number of fingers in images. Some are using these rules of thumb to accuse others of using GenAI. P3 recalled: ``I see AI-generated fan art come across my timeline way more often but usually it's coming across because somebody is quote retweeting the fan art and saying like `this is AI-generated and you can tell because of like this, this and this.''' At the same time, P9 doubted the efficacy of these strategies: ``I really wish it could always be true that we can tell when there's a human behind stuff but I think the reality of the situation is that we can't.'' As these folk theories are imperfect, artists now run the risk of being falsely accused of using AI: 
    
    \begin{quote}
        ``I have seen artists who have genuinely created this work get replies saying, `this is AI,' even when it's not so, it adds, I think, a level of frustration. That not only has have people had their work ripped off but now their work is being accused of not being theirs in the first place.'' (P10)
    \end{quote}
    
      Due to the risk of AI accusations, artists must sometimes \textit{prove} the authenticity of their work. P3 recalled: ``I have heard of my friends who are fan artists saying that now they try and draw in a very much more distinctive art style or when they draw they'll do time lapses of like them drawing the actual picture so they don't get accused of it being AI.''
      
      These strategies of detecting and calling out GenAI are an attempt for artists to regain agency, but these strategies also risk further disrupting our participants' artistic communities. In addition to the added labor of collectively authenticating others' work, artists now risk being accused of using GenAI themselves. In response, some are reconfiguring their art-making processes to prepare to defend against GenAI accusations. 

\subsection{The Queer Aesthetic Potential of GenAI}
\label{sec:benefit}

Queer aesthetic sensibilities often challenges dominant norms, upending conventional notions of ``good'' or ``bad'' art \cite{munoz2019cruising}. One may find queer beauty in, for instance, a film so earnestly bad that it becomes good \cite{sontag2018notes}. Likewise, Muñoz highlights the queer joy in strange or avant-garde forms, such as queer performance art or postmodern dance. Across these examples one can also see an affinity with failure. For Muñoz, queer failure ``helps the spectator exit from the stale and static lifeworld dominated by the alienation, exploitation, and drudgery associated with capitalism'' \cite[p. 173]{munoz2019cruising}. In a similar vein, our participants found the most queer artistic potential in the failures, strangeness, and glitches of GenAI.

    Participants often found GenAI most artistically interesting when artists lean into the nonhuman and surreal aspects of the technology, rather than trying to replace human creativity. P8 was most interested in how GenAI can create a ``picture of something that doesn't exist and can't possibly exist in the real world,'' which ``harkens back to [surrealist] art.'' Similarly, P9 enjoyed the ``early days of AI-generated images'' because they were ``pretty freaky.'' P9 elaborated:
    
    \begin{quote}
    
        ``[Early AI art was] way more art than what the AI image generators are spitting out now because you could see a machine struggling to comprehend stuff. And I that that was pretty cool because it wasn't really comparable to what I think that artists could make. So it was like its own niche. But instead of becoming its own niche it's now intruding in artist territory, and I think that's bad.'' (P9)
        
    \end{quote}

    Praising earlier uses of AI to create ``non-representational'' art, what P14 found most interesting about GenAI in the context of art ``is that [GenAI] is so non-human.'' He went on to explain: ``I could probably find hypothetical ways that I would use AI in a project and make it not corny, but I think it's very hard to do that right now because our imagination around AI is collectively limited because of the way it's hyped. It's hyped as this product that can substitute for human creativity, which is obviously disrespectful to human creativity and also to the particular creativity that AI offers.'' 

    Like P14, P6 was excited about the potential applications of GenAI in ``data bending'' glitch art, which he described as ``some of the most surreal — inhuman in a good way — feeling art.'' In particular, P6 was excited about the use of data poisoning tools, noting that ``most people'' are using data poisoning tools to ``mess with the datasets so [developers] can't use [these image models] anymore.'' However, P6 thought ``it's cool that the datasets are getting messed with because now we are going to get something different out of those generative image AI. We're going to get something that we wouldn't expect.'' For some, data poisoning is a way to refuse participating in the development of GenAI models. However, P6 finds excitement in datasets ``getting messed with'' because it may lead to unexpected model behaviors. Rather than a refusal of GenAI altogether, P6 finds pleasure in the queer refusal of dominant GenAI aesthetics in favor of ``something different.''

    As we described in Section \ref{sec:harm}, our participants were critical of those using GenAI to replace human creativity or produce art more quickly. At the same time, some saw the ``freaky'' (P9), ``surreal'' (P6, P8) and ``non-human'' (P14) styles possible with generative image models offering queer aesthetic possibilities. However, the development of increasingly realistic GenAI models both undermines the surreal aesthetic potentials of the technology and competes with the labor of human artists.

\section{Discussion}

    Contemporary narratives frame GenAI as inevitable---an industrial revolution, a frictionless engine of productivity \cite{ai_industrial_revolution}. But our participants enacted a different temporality. By refusing to use GenAI or removing their art from searchable platforms, they engage in what we might call queer refusals of ``straight time'' \cite{munoz2019cruising}, resisting the linear logics of innovation, efficiency, and profit-maximization. Of course, these refusals are not necessarily unique to those who identify \textit{as} queer artists. Rather, queer theory draws attention to those who challenge dominant norms. Our participants' refusals recall the politics of the Luddites not as anti-technology reactionaries, but as workers who recognized that the future on offer was not meant for them \cite{merchant_king_luddite, sabie2023unmaking, gerbracht2024we, reyes2025resisting}. Moreover, our work demonstrates how non-use can help individuals reclaim agency \cite{baumer2018departing, li2019protest_users} and engage in collective resistance \cite{vincent2021data}.
    
    In this section, we first discuss how CSCW researchers can help people, including but not limited to queer artists, refuse to participate in GenAI development. Then, we consider how the design of online communities might help people resist AI-generated media. At the same time, Muñoz suggests that queer refusal involves more than a rejection of the present ``repressive social order,'' such as cis-heteronormative and capitalist exploitation. This queer refusal is coupled with an ``insistence on potentiality or concrete possibility for another world.'' Queerness, he writes, “is not yet here,” but it gives us “a sense of the warm illumination of a horizon imbued with potentiality.” In our final discussion sections, we explore how researchers might support queer artistic futures beyond capitalist AI imaginaries and, more broadly, support artistic labor.
    
\subsection{Resisting Non-Consensual Generative AI Development}

    Our findings complicate assumptions that inclusion in GenAI datasets is inherently beneficial. Erasure is only harmful when one \textit{wants} to be included \cite{light2011hci}. While prior work critiques the inadvertent exclusion of queer people from training data \cite{dodge2021documenting}, many of our participants explicitly \textit{did not} want their art included. Echoing Muñoz’s critique of assimilationism and critical computing scholarship on refusal \cite{hoffmann2021terms,haimson2025trans}, simply “fixing” GenAI by adding more queer art does not support queer artists. In line with calls to move beyond deficit-centered design \cite{cunningham2023grounds,to_everyday_dis_2023,taylor2024carefully}, we argue that CSCW should instead focus on users who seek to \textbf{refuse} non-consensual scraping.

    One approach is improving transparency into what has already been taken. For instance, researchers might audit and make visible the content in prominent open-source datasets, such as prior investigations of the Book3 \cite{reisner2023revealed} dataset, or design tools that let artists check if their work was scraped \cite{reisner2023these}. Yet, as prior CSCW work on dataset traceability shows, tracking sources is difficult and incomplete \cite{scheuerman2023human}. Moreover, because companies rarely disclose training data, retroactive audits alone are insufficient. There is also a need for proactive mechanisms to prevent non-consensual scraping.

     As shown in recent CSCW research on non-consensual intimate media, designing for consent requires thinking across the many layers of infrastructure undergirding online media circulation \cite{qiwei2024sociotechnical}. At an infrastructural level, technologists could update internet standards — such as robots.txt — to differentiate between scraping permissions for search engine indexing versus GenAI training \cite{jimenez2024ai}. However, such changes may prove ineffective because internet standards are usually enforced by norms rather than laws \cite{lessig_code_1999}. At the application level, developers might imperceptibly perturb images and videos that users upload \cite{shan2023glaze} to protect these works from being used to build GenAI models, similar to social media sites removing location metadata when users upload images \cite{geeng2020usable, henne2013snapme}. At the interface level, platforms should provide clear, default opt-outs, while at the policy level researchers can advocate to extend ``right to be forgotten'' protections to training datasets \cite{novelli2024generative}.
     
     It is unlikely that for-profit companies will make decisions that are best for users at the expense of their bottom-lines, as indicated by Reddit and Tumblr deciding to sell user data to GenAI developers \cite{Ohlheiser_2024}. Moreover, as indicated by the numerous lawsuits facing tech companies at the time of our writing over GenAI scraping practices \cite{Jahner_2024, Robertson_2024}, tech companies have a long history of skirting regulation by ``asking for forgiveness rather than permission'' \cite{vaughan2024new}. There will always be tech companies and individuals that choose to ignore artists' wishes. Technologists have spent decades developing DRM (Digital Rights Management) software to prevent everyday users from scraping artwork owned by major corporations \cite{doctorow2014information}, with little equivalent effort to safeguard everyday artists \cite{lessig2008remix}. To reverse this course, CSCW might more actively pursue oppositional privacy research, developing and coordinating protective tools that give artists collective leverage against non-consensual generative AI development \cite{shan2023glaze, vincent2021data,shan2024nightshade}.

\subsection{Resisting AI-Generated Media in Online Communities}

Our participants described online artistic communities as spaces where they came to understand themselves, connect with others, and form queer communities. However, these spaces have been profoundly unsettled by fears of non-consensual GenAI training and the circulation of undisclosed AI-generated art. Muñoz reminds us that queer world-building is not about securing safety within the present but about reaching toward alternative futures---what he calls a stepping out of ``this place and time to something fuller, vaster, more sensual, and brighter'' \cite{munoz2019cruising}. From this perspective, anti-GenAI communities are not merely defensive responses but utopian projects. They reject extractive platform logics and imagine worlds otherwise in which artistic labor, intimacy, and collectivity can thrive on different terms.

This utopian lens shifts how we might understand emerging anti-AI social media platforms, like Cara \cite{cara_anti_ai}. Rather than viewing them only as replacements for compromised spaces like DeviantArt \cite{deviant_art_sellout}, we can see them as enactments of queer disidentification---forms of refusal that carve out futures resistant to commodification. Paralleling CSCW research on fan migration and platform building \cite{dym2022building, fiesler2020moving, fiesler2016archive}, these communities are performing the very work of world-making that Muñoz theorizes, building infrastructures that insist on collective flourishing outside the inevitability of GenAI. Future research might ask how to design with and for such resistant communities, not simply to mitigate harms but to cultivate their utopian horizons.

Participants also described developing ``folk theories'' to detect undisclosed GenAI art. At the same time, this heightened suspicion places new burdens on artists and audiences to expend additional effort to prove or disprove AI involvement when sharing or enjoying art. These everyday audits \cite{shen2021everyday} highlight both the creativity and the burdens of grassroots governance. Similarly, participants expressed interest in labeling GenAI-made art when shared on social media, seeing such labels as a way to resist GenAI by filtering out AI-generated content. At the same time, implementing labeling raises difficult design challenges. Because participants' perceptions of art depended on precisely where GenAI was used in the creative process, a single ``AI-generated'' label risks oversimplifying the diverse ways GenAI may be involved in art-making. Future work might therefore explore labeling schemes that move beyond a simple binary of GenAI use/non-use \cite{epstein2023label} as well as examine how different interface designs for displaying labels shape audience perceptions \cite{habib2020s}. Complementary lines of research could continue investigating watermarking or detection methods for AI-generated media \cite{zhong2023copyright, knott2023generative}, while also weighing the relative harms of false-negatives versus false-positives across application domains \cite{dalalah2023false}.

Muñoz's queer theory cautions us not to naturalize these authentication and labeling practices as endpoints of design. Much like marginalized artists have critiqued GenAI moderation systems in prior work \cite{mirowski2024robot, taylor2025straightening}, systems that institutionalize collective labeling or forced authentication risks shifting power toward technology developers and away from artists. Compulsory verification undermines artists' ability to embrace opacity or refuse disclosure. Thus, future research should explore not only technical methods for authentication, watermarking, and labeling \cite{zhong2023copyright, knott2023generative, epstein2023label}, but also how to design with refusal in mind. While some communities may collectively value provenance, others may resist forced authentication as forms of surveillance. To follow Muñoz is to treat refusal itself as world-building: enabling artists to say no to GenAI and no to compulsory verification, as part of imagining futures otherwise. In this way, resistance to AI-generated media is not just about preventing harm in the present but about opening possibilities for queer worlds not yet here.

\subsection{Supporting Queer Artists}

    For Muñoz, queer art entails a refusal of capitalist logics of productivity and profit-maximization, instead cultivating alternative temporalities and aesthetics that gesture toward utopian futures  \cite{munoz2019cruising}---an orientation echoed by our participants. While they were highly critical of managers using GenAI to cut production costs (Section \ref{sec:harm}), participants expressed fewer concerns when artists used GenAI outside of profit-driven contexts (Section \ref{sec:benefit}). Building on these insights, we suggest that designers can support queer futures by pursuing alternative paradigms that decenter efficiency and commodification.
    
    One such paradigm to counter ``straight'' temporalities is slowness \cite{odom2014slow, odom2018slow, odom2015understanding}. In response to participants' critiques of GenAI being used to sacrifice quality for efficiency, models that emphasize slow output could open space for reflection and resist managerial appropriation. For example, a system that produces only one image per week could invite more deliberate engagements with GenAI's artistic potential, while also reducing the environmental costs of large-scale AI development \cite{dodge2022measuring}. Such approaches align with post-growth HCI calls for technologies oriented around sustainability and care rather than acceleration \cite{sharma2023post_growth, sharma2017analyzing}.

    Researchers could also explore legal mechanisms to resist commodification. For example, “copyleft” licenses, developed in free and open-source communities, prevent enclosure by requiring derivative works to remain free \cite{soderberg2002copyleft}. Future computing research could collaborate with legal scholars to design similar licenses for GenAI models and datasets. Such GenAI models would allow artists to engage in minor support work, like practicing techniques, but may be unappealing to for-profit companies.

    Finally, our participants underscored the importance of imagining queer artistic futures beyond GenAI altogether. Echoing accessibility research that highlights relational support over individual productivity \cite{bennett2018interdependence}, we suggest shifting design attention toward infrastructures that sustain queer artistic mutual aid. In other words, researchers can design ways for queer artists to collectively support one another without having to rely on GenAI. For instance, systems could facilitate gifting and appreciation practices among queer artists, extending prior work on recognition in open-source communities \cite{khadpe2025hug}. P3, for example, described coordinating with others to create zines supporting queer and disabled causes, pointing toward broader design possibilities for technologies that scaffold collective art-making, as in Allred \& Gray’s work on activist fanzines \cite{allred2021gay}. While GenAI is increasingly central in HCI research \cite{pang2025understanding}, our participants did not necessarily seek ``better'' models. Instead, their practices invite CSCW scholars to imagine queer artistic futures beyond GenAI by centering queer relationality.

\subsection{Supporting Queer Aesthetics}

    Muñoz emphasizes queer art's subversive aesthetics and embrace of strangeness \cite{munoz2019cruising}. Likewise, some of our participants enjoyed the weird or surreal style of early GenAI models, while dismissing new ones as  ``corny'' simulacra of human creativity due to increasing realism. By praising the surrealism of older GenAI models, our findings echo prior work on the preferences for glitches among some artists using visual GenAI \cite{sivertsen2024machine, caramiaux2022explorers, chang2023prompt, taylor2025straightening}. Artists' dissatisfaction with the style of newer models calls into question what it means for a GenAI model to be ``state-of-the-art'' for artists.
    
    Researchers are increasingly focusing on how to improve the evaluation or measurement of GenAI model capabilities \cite{wallach2025position}. In light of our participants' preferences for the surreal style of older generative image models, future research should study how AI developers are measuring the aesthetic quality of images. For example, \citet{taylor2026algorithmicgazeimagequality} recently investigated the LAION-Aesthetics Predictor, a visual aesthetic evaluation model commonly used to curate images for GenAI training data. Paralleling the biases toward realism our participants critiqued, the authors found that the LAION-Aesthetics Predictor rates realistic art more highly than abstract or surreal art. In addition to studying these technical infrastructures, we encourage CSCW researchers to investigate the human infrastructure \cite{lee2006human_infra} undergirding aesthetic evaluation, such as the data labor of those who label the aesthetic quality of images.

    Designing generative AI models \textit{for} queer aesthetics may also support more ethical engagements with artists. As noted in prior work on content moderation, \textit{scale} is commonly used as an excuse for technology developers to avoid accountability \cite{gillespie2020content}. According to our participants, the prevailing paradigm of training generative AI models through large-scale data collection without consent \cite{schuhmann2022laion} has immensely harmed queer artistic communities. However, one may not need as much data to create glitchy or surreal generative image models, such as Google's 2015 model DeepDream \cite{mordvintsev2015inceptionism}. Therefore, one may be able to design generative image models \textit{for} queer aesthetics using smaller datasets collected with creators' consent. In service of these efforts, we encourage researchers to explore the collaborative construction of consensual datasets as a CSCW problem, such as the community-based development of the World Wide Dishes text-to-image evaluation dataset \cite{magomere2025world}.

\subsection{Supporting Artistic Labor}

    Our participants drew a sharp line between GenAI as peripheral aid—such as editing pitch-letters—and GenAI as replacement for the core labor of art-making. This distinction exposes a broader politics of invisibility. As Becker notes, art worlds depend on ``support work'' that is usually erased, yet without which art cannot exist \cite{becker2023art}.  GenAI risks deepening this erasure, framing support tasks as disposable and as interchangeable with machines.

    Muñoz reminds us that queer practices of labor are not simply about producing commodities but about resisting the ``straight time'' of efficiency and profit. In this sense, what is at stake is not just the preservation of jobs, but the defense of queer temporalities---slowness, care, collaboration---that capitalist logics render unproductive. The imposition of GenAI into creative labor markets exemplifies what Muñoz critiques as assimilationist compromise, demanding that artists adjust to extractive systems rather than reimagining them.
    
    Rather than treating GenAI as an inevitable supplement to creative work, CSCW researchers must confront the structures that naturalize dispossession. This might mean aligning with unions and creative collectives not only to negotiate contract language but to contest the very framing of artistic labor as fungible, flexible, or optional. GenAI's appropriation of queer art and labor is not simply a technological shift. It is a continuation of a long history of capitalist exploitation of marginalized creativity. Refusing this future requires imagining infrastructures that do not simply ``accommodate'' GenAI but actively block its coercive uptake, enabling artists to sustain labor practices grounded in solidarity, opacity, and utopian potential rather than extraction \cite{greenbaum1996back, tang2023back_to_labor}.

\subsection{Limitations}

    Our work does not intend to represent the views of all ``queer artists.'' Instead, we examine the views of a particular group of queer artists, which we interpret using José Esteban Muñoz's theorization of queer aesthetics. Notably, all our participants lived in the United States of America, and our interviews were conducted in English. Owing to the cultural situatedness of gender and sexuality \cite{nova2021facebook} and artistic practices \cite{mim2024between}, more work is needed to understand the queer artists' experiences beyond the US. 
    
    Our findings also do not fully represent the diversity of queer artistic experiences \textit{within} the US. As our interviews were conducted over Zoom, participants needed a reliable internet connection in order to participate. Also, Black queer artists have been pivotal to queer art, but are not represented in our study \cite{munoz2019cruising, smalls2019introduction}. This may be due to a limitation in having only recruited from Twitter and Reddit, as fewer Black people in the US use these sites than Instagram or TikTok \cite{pew_social_media}. Our participants also skewed young, with a median age of 26 and our oldest participant being 33. As noted in prior Queer HCI scholarship, there is a need for future research on the experiences of older adults \cite{zhu2025challenges, taylor2024cruising}.
    
    While in this work we recruited self-identified queer artists, not every LGBTQ+ person would necessarily identify as queer \cite{taylor2024cruising}. As Muñoz notes, ``queerness'' is historically associated not only to marginalized genders and sexualities but also a form of radical politics and non-normative aesthetics \cite{munoz2019cruising}. Therefore, our findings may have differed if we instead recruited LGBTQ+ identifying artists. Instead of representing LGBTQ+ artists writ large, our findings are likely more transferable to artistic communities that similarly value relationality and anti-capitalism, such as the transformative fan community to which at least 6 of our participants belonged. At the same time, an important limitation of our work is that, although our participants felt strong solidarity with working artists, most of our participants did not make a living primarily through their art. As some of our participants engaged in artistic communities that stigmatize profiting from one's work \cite{fiesler2014remixers}, our participants should not be disregarded as merely amateur artists. That said, the attitudes of, for instance, full-time screenwriters toward GenAI may differ from those of our participants.

    As transgender, non-binary and disabled people in the US tend to view AI more negatively than members of dominant social groups \cite{haimson2025ai_attitudes}, our participants' negative attitudes toward and non-use of GenAI are to be expected. At the same time, this non-use may have shaped our participants' attitudes towards GenAI. This may be, for instance, why LGBTQ+ biases in model behaviors did not figure prominently in our findings. Moreover, our participants' conceptions of GenAI tended to focus on prompt-based interaction paradigms, such as text-to-image models and chatbots. In light of our findings that participants were more comfortable with AI being used for support work, future research should study the acceptability of other types of GenAI models, such as those used for coloring. Additionally, Lucy Suchman notes that the category of technologies referred to as ``AI'' is quite expansive \cite{suchman2023uncontroversial}. As a result, artists likely have different conceptions of what constitutes AI and its acceptability. Similar to prior research on people's mental models of the internet \cite{kang2015mental_models}, future research should explore artists mental models of AI and AI development. Finally, while this work largely focuses on queer artists refusing GenAI, we encourage future research into queer artists who have chosen to integrate GenAI in their artistic practices \cite{taylor2025straightening}.

\section{Conclusion}

In this work, we examined how queer artists are making sense of and responding to the recent proliferation of GenAI. The tensions our participants identified between themselves and GenAI are not simply the most recent version of longstanding aesthetic debates \cite{becker2023art} or the result of technological \textit{naïvetés}. Instead, these are tensions between competing future visions of what art is and should be, between art-making as a relational queer practice versus an alienated commodity from which to extract profit. We encourage CSCW researchers to contest these dominant AI imaginaries. As José Esteban Muñoz reminds us, queerness demonstrates the ``concrete possibility for another world'' \cite{munoz2019cruising}.

\bibliographystyle{ACM-Reference-Format}
\bibliography{refs}

\begin{acks}
We thank Amy Bruckman, Rose Chang, Mary Gray, Anna Fang, Oliver Haimson, Shivani Kapania, Franchesca Spektor, Cella Sum, Sherry Tongshuang Wu and our reviewers for their feedback on this work. This research was supported, in part, by the National Science Foundation (award number 2442153).
\end{acks}

\end{document}